\documentstyle[prd,aps,eqsecnum,epsf,array]{revtex}

\newcommand{\be}{\begin{equation}}
\newcommand{\ee}{\end{equation}}

\newcommand{\bqn}{\begin{eqnarray}}
\newcommand{\eqn}{\end{eqnarray}}

\begin{document}

\setlength{\topmargin}{-2ex}

\draft

\title{Time evolution  of quantum systems in microcavities and in free space -- 
a non-perturbative approach}
\author{G. Flores-Hidalgo$^\dag$, A. P. C. Malbouisson$^\star$}

\address{$^\dag$ Instituto de F\'{\i}sica Teorica, UNESP, Rua Pamplona 145,
01405-900, São Paulo, SP, Brazil\\
$^\star$CBPF/MCT, Rua Dr.
Xavier Sigaud 150, Urca,
22290-180, Rio de Janeiro, RJ, Brazil}

\maketitle

\begin{abstract}
We consider a system consisting of a particle in the harmonic approximation,
having frequency $\bar{\omega}$, coupled to a scalar field inside a
spherical reflecting cavity of diameter $L$. By introducing {\it dressed} coordinates
we define {\it dressed} states which allow a non-perturbative unified description of the 
radiation process, in both cases, of a finite or an arbitrarily large cavity, for  weak
and strong coupling regimes. In the case of weak coupling, we recover from our exact
expressions the well known decay formulas from perturbation theory. We perform a study of the energy distribution in a small cavity, with the initial condition 
that the particle is in the first excited state. In the {\it weak} 
coupling regime, 
we conclude for the quasi-stability of the 
excited particle. For instance, for a frequency $\bar{\omega}$ of the order $\bar{\omega}\sim 
4.00\times 10^{14}/s$ (in the visible red), starting from the initial condition that the 
particle is in the first excited level, we find that for a cavity with diameter $L\sim 1.0
\times 10^{-6}m$, the probability that the particle be at any time still in the first excited 
level, will be of the order of $97\%$. The behaviour of the system for {\it strong} coupling is rather different from 
its behaviour in the {\it weak} coupling regime. For appropriate 
cavity dimensions, which are of the same order  of those ensuring stability for weak 
coupling, we ensure for strong coupling the complete decay of the particle to the 
ground state in a small ellapsed time.
 Also we consider 
briefly the effects of a quartic interaction up to first order in the interaction parameter 
$\lambda$. We obtain for a large cavity an explicit $\lambda$-dependent expression for the 
particle radiation process. This formula is obtained in terms of the corresponding exact
expression for the linear case and we conclude for the enhancement of the particle decay 
induced by the quartic interaction.
\end{abstract}

\section{Introduction}

Exact solutions of problems in theoretical physics is known to researchers since a long 
time ago to be a rather rare situation. It is a common feature to different branches of 
physical sciences, such as celestial mechanics, field theory and statistical physics, 
that exact solutions of coupled equations describing the physics of interacting bodies 
is a very hard problem. In statistical physics and constructive field theory,
general theorems can be derived using cluster-like expansions and other related methods 
\cite{Jaffe}. In some cases, these methods allow the rigorous construction of field 
theoretical models (see for instance\cite{Jarrao} and other references therein),
but, in spite of the rigor and in some cases the beauty of demonstrations, these methods 
do not furnish useful tools for calculations of a predictive character.
Actually, apart from computer calculations in lattice field theory, the only available 
method to solve this kind of problems, except for  a few special cases, is perturbation 
theory. In modern physics, a prototype situation for instance  in abelian gauge theories, 
is a system composed of a charged body described by a matter field interacting with a neutral
(gauge) field through some (in general non linear) coupling characterized by some parameter
$g$, usually named the coupling constant or the charge of the body. The perturbative
solution to this situation is obtained by means of the introduction of bare, non interacting 
matter and gauge fields, to which are associated bare quanta, the interaction being introduced 
order by order in powers of the coupling constant in the perturbative expansion for the 
observables. This method gives remarkably accurate results in Quantum Electrodynamics and in 
Weak interactions. In high energy physics, asymptotic freedom allows to apply Quantum 
Chromodynamics in its perturbative form and very important results have been obtained in this
way in the last decades.

In spite of its wide applicability, there are situations where the use of perturbation theory 
is not possible, as in the low energy domain of Quantum Chromodynamics, where confinement of 
quarks and gluons is believed to take place. 
In this particular situation, no analytical approach in the 
context of Quantum Field Theory is available up to the present moment. Also there are 
situations in the scope of Quantum Electrodynamics, in the domain of Atomic Physics, Cavity 
Quantum Electrodynamics and Quantum Optics, where perturbation methods are of little usefulness.
The theoretical understanding of these effects on perturbative grounds requires the 
calculation of very high-order terms in perturbation series, what makes standard Feynman
diagrams technique practically unreliable in those cases \cite{bouquinCohen}. In this article 
we study situations where these mathematical difficulties can be circumvected.
We consider systems that under certain conditions may be approximated by the system composed 
of a harmonic oscillator coupled {\it linearly} to the modes $\omega_{i}$ of a scalar field 
trough some effective coupling constant $g(\Xi)$, the whole system being confined in 
a cavity $\Xi$. We present an unified formalism for the truly confined system in a small cavity 
and for the system in free space, understood as the limit of a very large cavity. Linear 
approximations are used in several contexts, as in the general QED linear response theory, 
where the electric dipole interaction gives the leading contribution to the radiation process
\cite{McLachlan,Jhe} In cavity QED, in particular in the theoretical investigation of 
higher-generation Schrodinger cat-states in high-Q cavities, a linear model is employed 
\cite{Jmario}. Also, approaches of this type have been used in condensed matter physics for 
studies of Brownian motion \cite{zurek,paz} and in quantum optics to study decoherence, by 
assuming a linear coupling between a cavity harmonic mode and a thermal bath of oscillators at 
zero temperature \cite{Davidovitch,Fonseca}.
In another mathematical framework, there are a large number of successful attempts in the 
literature to by-pass the limitations of perturbation theory. In particular, there are methods to 
perform resummations of perturbative series (even if they are divergent), which ammounts in 
some cases to analytically continue weak-coupling series to a strong-coupling domain
\cite{Le Guillou,Weniger,Jents,Cvetic,ari1,simao,ari2,adolfoso}. For instance, starting from
a function of a coupling constant $g$ defined formally by means of a series (not necessarily
convergent),
\be
f(g)=\sum_{n^=0}^{\infty}a_{n}g^{n}\;,
\label{f}
\ee
we can, under certain analyticity assumptions (the validity of the Watson-Nevanlinna-Sokal
theorem, see for instance \cite{Rivasseau} and other references therein), define its
Borel transform as the associated series,
\be
B(b)=\sum_{n^=0}^{\infty}\frac{a_{n}}{n!}b^{n}\;,
\label{b}
\ee
which has an analytic continuation on a strip along the real $b$-axis from zero to infinity.
It can be easily verified that the inverse Borel transform
\be
\tilde{B}(g)=\frac{1}{g}\int_{0}^{\infty}db\;e^{b/g}B(b)\;,
\label{invb}
\ee
reproduces formally the original series (\ref{f}). From a physical point of view, the
important remark is that the series $B(b)$ can be convergent and summed up even if the
series (\ref{f}) diverges. In this case, the inverse Borel tranform (\ref{invb}) defines a 
{\it function} of $g$, $\tilde{B}(g)$, which we can think about as the "sum" of the divergent 
series (\ref{f}). This function $\tilde{B}(g)$ can be defined for values of $g$ not necessarily 
small and, in this sense we can perform an analytic continuation from a weak to a 
strong-coupling regime. Techniques of this type are of a predictive character and have been 
largely employed in the last years in Quantum Field Theory literature.

Nevertheless as a matter of principle, due to the non vanishing of the coupling constant, 
the idea of a bare particle associated to a bare matter field is actually an artifact of 
perturbation theory and is physically meaningless. A charged physical particle is always 
coupled to the gauge field, in other words, it is always "dressed" by a cloud of quanta of 
the gauge field (photons, in the case of Electromagnetic field). In fact as mentioned above,
from a phenomenological point of view there are situations even in the scope of QED, where
perturbation methods are of little usefulness, for instance, resonant effects associated to 
the coupling of atoms with strong radiofrequency fields. As remarked in \cite{bouquinCohen}, 
the theoretical understanding of these effects using perturbative methods requires the 
calculation of very high-order terms in perturbation theory, what makes the standard Feynman
diagrams technique practically unreliable. The trials of treating non-perturbativelly systems 
of this type, have lead to the idea of "dressed atom", introduced originally in Refs.
\cite{Polonsky} and \cite{Haroche}. Since then this concept has been  used  to investigate
several situations involving the interaction of atoms and electromagnetic fields, as for 
instance, atoms embedded in a strong radiofrequency field background  and atoms in intense 
resonant laser beans.

In order to give a precise mathematical definition and a clear physical meaning to the idea 
of dressed atom, we do not face the non-linear character of the problem involved in realistic
situations, we consider instead, as mentioned above a linear problem, in which case a 
rigorous definition of "dressed atom" and more generally of a dressed particle interacting
weakly or strongly with the field 
can be given. In other words, in this paper we adopt 
a general physicist's point of view. We do not intend to describe all the specific features of 
a real non-linear physical situation. Instead we analyse a simplified linear version of the 
atom-field or particle-environment system and we try to extract the most detailed information 
we can from this model. We will introduce {\it dressed} states, by means of a precise and 
rigorous definition to solve our problem. Our {\it dressed} states can be viewed as a rigorous 
version of the semiqualitative idea of dressed atom mentioned above, which can be constructed 
in reason of the linear character of our problem. Our aim in taking  a linear model is to
have a clearer understanding of what is one of the essential points, namely, the need for
non-perturbative analytical treatments of coupled systems, which is the basic problem underlying
the idea of a {\it dressed} quantum mechanical system. Of course, such an approach to a 
realistic non-linear system is an extremely hard task, and here we achieve what we think is a 
good agreement between physical reality and mathematical reliability.

In recent publications \cite{adolfo1,adolfo2} a method ({\it dressed} coordinates and 
{\it dressed} states) has been introduced that allows a  non-perturbative approach to situations 
of the type described above, provided that the interaction between the parts of the system can 
be approximated by a linear coupling. More precisely, the method applies for all systems that 
can be described by an Hamiltonian of the form,

\begin{equation}
H=\frac{1}{2}\left[p_{0}^{2}+\omega_{0}^{2}q_{0}^{2}+
\sum_{k=1}^{N}(p_{k}^{2}+\omega_{k}^{2}q_{k}^{2})\right]-q_{0}\sum_{k=1}^{N}c_{k}q_{k},
\label{Hamiltoniana}
\end{equation}

where the subscript $0$ refers to the particle and $k=1,2,...N$ refer to the harmonic 
environment modes. The limit $N\rightarrow \infty$ in Eq.(\ref{Hamiltoniana}) is understood.
The equations of motion are,

\begin{equation}
\ddot{q}_{0}(t)+\omega_{0}^{2}q_{0}(t)=\sum_{k=1}^{\infty}
c_{k}q_{k}(t)  
\label{eq. mov3.}
\end{equation}

and

\begin{equation}
\ddot{q}_{k}(t)+\omega_{k}^{2}q_{k}(t)=c_{k}q_{0}(t)\;.
\label{eq. mov4.}
\end{equation}

A Hamiltonian of the type above, describing a linear coupling of a particle with an environment, 
has been used in Refs. \cite{zurek,paz,ullersma,haake,caldeira,shram} to study the quantum Brownian
motion of a particle.  
In the case of the coupled atom field system, this formalism recovers the experimental observation 
that excited states of atoms (weakly) coupled to the electromagnetic field 
in sufficiently small cavities are stable. It allows to give formulas 
for the probability of an atom to remain excited for an infinitely long time, provided it is
placed in a cavity of appropriate size \cite{adolfo2}. For an emission frequency
in the visible red, the size of such cavity is in agreement with experimental
observations \cite{Hulet,Haroche3}.  The model that we consider for the particle-field system
consists of an one dimensional oscillator interacting with a massless scalar field and described by
the Lagrangean:

\be
{\mathcal L}=\frac{1}{2}\dot{q}_0^2-\frac{\omega_0^2}{2}q_0^2+\frac{1}{2}\int d^3\vec{x}~\!
\frac{1}{2}\partial_\mu\phi\partial^\mu\phi+ 2\pi\sqrt{g}\int d^3\vec{x}\phi(x) 
\delta(\vec{x})q_0\;,
\label{la1}
\ee

where $g$ is a coupling constant with dimension of frequency.

Notice that we are supposing that the particle interacts with the field by a contact term 
only at the origin or, in other words the particle is centered at $\vec{x}=0$. This is 
equivalent to the dipole approximation for electromagnetic interactions.
 Next we confine the whole system into a sphere 
of diameter $L$. Writing
\be
\phi(\vec{x},t)=\sum_kq_k(t)\phi_k(\vec{x})\;,
\label{la2}
\ee
where the $\phi_k(\vec{x})$ form an orthonormal basis obeying the equation,
\be
-\vec{\nabla}^2\phi_k(\vec{x})=\omega_k^2\phi_k(\vec{x})\;,
\label{la3}
\ee
replacing Eq. (\ref{la2}) in Eq. (\ref{la1}) and using Eq. (\ref{la3}) we obtain
\be
{\mathcal L}=
\frac{1}{2}\left[\dot{q}_0^2-\omega_0^2q_0^2+\sum_k(\dot{q}_k^2-\omega_k^2q_k^2) \right]
+  2\pi\sqrt{g}q_0\sum_k\phi_k(\vec{0})q_k\;.
\label{la4}
\ee

Notice that the field modes that interact with the particle are those modes such that
$\phi(\vec{0})\neq 0$.
These modes, given by the eigenvalues equation (\ref{la3}), possess spherical
symmetry.  Then, solving the eigenvalues equation (\ref{la3}) in spherical coordinates, with
the  boundary condition $\phi_k(L/2)=0$, we get for the spherically symmetric modes
\be
\phi_k(\vec{x})=\frac{\sin(\omega_k r)}{\sqrt{\pi L}~\!r}\;,~~~\omega_k=\frac{2\pi k}{L}\;,
~~~k=1,2,3,...,
\label{la5}
\ee
where $r=|\vec{x}|$. From the Lagrangean (\ref{la4}) we can get the Hamiltonian
in a standard way. Using Eq. (\ref{la5}) we obtain the Hamiltonian  given by Eq.
(\ref{Hamiltoniana}) with the coupling coefficients $c_k=\eta\omega_k$, where
$\eta=\sqrt{2g\Delta\omega}$ and
$\Delta\omega$ is the interval between two neighbouring bath (field) frequencies.

\section{The eigenfrequencies spectrum and the diagonalizing matrix}
The bilinear Hamiltonian (\ref{Hamiltoniana}) can be turned to principal axis by means of a
point
transformation,
\be
q_{\mu}=\sum_{r=0}^Nt_{\mu}^{r}Q_{r}\;,~~~p_{\mu}=\sum_{r=0}^Nt_{\mu}^{r}P_{r}\;,
~~~ \mu=(0,\{k\})\;,~~~k=1,2,..., N\;,
\label{transf}
\ee
performed by an orthonormal  matrix $T=(t_{\mu}^{r})$. The subscripts $\mu=0$ and $\mu=k$ refer 
respectively to the particle and to the harmonic modes  of the bath and $r$ refers to the normal 
modes. In terms of normal momenta and coordinates, the transformed Hamiltonian in principal axis 
reads,

\begin{equation}
H=\frac{1}{2}\sum_{r=0}^{N}(P_{r}^{2}+\Omega_{r}^{2}Q_{r}^{2})\;,
\label{diagonal}
\end{equation}
where the $\Omega_{r}$'s are the normal frequencies corresponding to the possible collective 
stable oscillation modes of the coupled system. Using the coordinate transformation
$q_{\mu}=t_{\mu}^{r}Q_{r}$ in the equations of motion and explicitly making use of the 
normalization condition $\sum_{\mu=0}^{N}(t_{\mu}^{r})^{2}=1$,  we get,
\begin{equation}
t_{k}^{r}=\frac{c_{k}}{(\omega_{k}^{2}-\Omega_{r}^{2})}t_{0}^{r}\;,
~~~
t_{0}^{r}= \left[1+\sum_{k=1}^{N}\frac{c_{k}^{2}}{
(\omega_{k}^{2}-\Omega_{r}^{2})^{2}}\right]^{-\frac{1}{2}}
\label{tkrg1}
\end{equation}
and the normal frequencies $\Omega_r$ are given as solutions of the equation,
\begin{equation}
\omega_{0}^{2}-\Omega_{r}^{2}=\sum_{k=1}^{N}\frac{c_{k}^{2}}
{\omega_{k}^{2} -\Omega_{r}^{2}}\;.
\label{Nelson1}
\end{equation}
Remembering $c_k=\eta\omega_k$, Eq. (\ref{Nelson1}) can be written as
\be
\omega_0^2-N\eta^2-\Omega_r^2=\eta^2\sum_{k=1}\frac{\Omega_r^2}
{\omega_k^2-\Omega_r^2}\;.
\label{la5b}
\ee

In the limit $N\rightarrow \infty$ Eq.(\ref{la5b}) is meaningless if $\omega_{0}$ is finite. To overcome this difficulty we define $\omega_0^2$ as containing
the divergent term $N\eta^2$ and a finite part,

\be
\omega_0^2=\bar{\omega}^2+N\eta^2\;,
\label{la6}
\ee
where $\bar{\omega}^2$ is finite and is defined as the physical (squared) frequency, 
while $\omega_0^2$ is not physical, it as a bare (squared) frequency and defined in such a way 
that the divergent terms in the left hand side of Eq. (\ref{la5b}) cancel. Replacing
Eq.(\ref{la6}) in Eq. (\ref{la5b}) we obtain
\be
\bar{\omega}^2-\Omega_r^2=\eta^2\sum_{k=1}^N\frac{\Omega_r^2}{\omega_k^2-\Omega_r^2}\;.
\label{la7}
\ee
We see that the above procedure is exactly the analogous of naive mass renormalization in
quantum field theory: the addition of the counterterm $-(\eta^2N/2) q_0^2$ allows to 
compensate the infinity of $\omega_0^2$ in such a way as to leave a finite, physically meaninful 
renormalized frequency $\bar{\omega}$. This simple renormalization scheme has been originally 
introduced in Ref.\cite{tirring}. For a nice discussion on the subjec the interested reader can see Ref.\,\cite{weiss}.
Using the formula,

\begin{equation}
\sum_{k=1}^{N}\frac{1}{(k^{2}-u^{2})}= \left[\frac{1}{2u^{2}}-\frac{\pi}{u}
{\rm cot}(\pi u)\right]\;,
\label{id4}
\end{equation}
Eq. (\ref{la7}) can be written in closed form,
\begin{equation}
\cot(\frac{L\Omega}{2})=\frac{\Omega}{\pi g}+\frac{2}
{L\Omega}(1-\frac{L\bar{\omega}^{2}}{2\pi g})\;.
\label{eigenfrequencies2}
\end{equation}
The solutions of Eq. (\ref{eigenfrequencies2}) with respect to  $\Omega$ give the
spectrum of eigenfrequencies $\Omega_{r}$ corresponding to the collective normal modes.
The transformation matrix elements $T=(t_\mu^r)$ are obtained in terms of the physically meaningful 
quantities $\Omega_{r}$, $\bar{\omega}$, after some rather long but straightforward manipulations.  
They read,
\begin{equation}
t_{0}^{r}=\frac{\eta \Omega_{r}}{\sqrt{(\Omega_{r}^{2}-\bar{
\omega}^{2})^{2}+\frac{\eta^{2}}{2}(3\Omega_{r}^{2}-\bar{\omega}^{2})+
\pi^{2}g^{2}\Omega_{r}^{2}}}\;,~~~
t_{k}^{r}=\frac{\eta\omega_{k}}{\omega_{k}^{2}-\Omega_{r}^{2}}t_{0}^{r}\;.
\label{t0r2}
\end{equation}

In free-space, that is, in the limit $L\to\infty$ we get for $t_0^r$,
\be
t_0^r=\lim_{\Delta\Omega\to 0}\frac{\sqrt{2g}\Omega\sqrt{\Delta\Omega}}
{\sqrt{(\Omega_r^2-\bar{\omega}^2)^2+\pi^2g^2\Omega^2}}
\label{la8}
\ee
and for $t_k^r$,
\be
t_k^r=\frac{2g\omega_k\Delta\omega}{(\omega_k+\Omega_r)(\omega_k-\Omega_r)}
\frac{\Omega_r}{\sqrt{(\Omega_r^2-\bar{\omega}^2)^2+\pi^2g^2\Omega^2}}\;,
\label{la9}
\ee
where we use the fact that in the limit $L\to\infty$, 
$\Delta\Omega\to\Delta\omega=2\pi/L$.

\section{Dressed states and the emission process in free space}

We define below some coordinates $q^{\prime}_{0}$, $q^{\prime}_{i}$ associated  
respectivelly to the 
{\it dressed} particle (atom) and to the {\it dressed} bath (field). These coordinates will reveal 
themselves to be appropriate to give an appealing non-perturbative description of the 
particle-field system. 
The normalized eigenstates of our system (eigenstates of the Hamiltonian in principal axis) 
can be written in terms of the normal coordinates,    
\begin{eqnarray} 
\langle Q|n_{0},n_{1},...;t\rangle&\equiv &\phi_{n_{0}n_{1}...}(Q,t)\nonumber\\
&=&\prod_{s=0}^N\left[\sqrt{\frac{2^{n_s}}{n_s!}}H_{n_{s}}(\sqrt{\Omega_{s}}Q_{s})\right]
\Gamma_{0}(Q){\rm e}^{-iE_{n_0n_1...}t}, 
\label{autofuncoes} 
\end{eqnarray} 
where $H_{n_{s}}$ stands for the $n_{s}$-th Hermite polynomial and $\Gamma_{0}(Q)$ is the 
normalized ground state eigenfunction. The eigenvalues $E_{n_0n_1...}$ 
are given by
\be
E_{n_0n_1...}=\sum_{s=0}^N\left(n_s+\frac{1}{2}\right)\Omega_s\;.
\label{autoe}
\ee

Next we intend to divide the system into the {\it dressed} particle and the {\it dressed}
environment (bath, field) by means of some conveniently chosen {\it dressed} coordinates, $q_0'$ and
$q_j'$ associated respectively to the {\it dressed} particle and to the {\it dressed} 
oscillators composing the environment. These coordinates will allow  a natural
division of the system as composed of the {\rm dressed} (physically observed) particle and 
 the {\it dressed} environment. The {\rm dressed} particle will contain automatically
all the effects of the environment on it. Clearly, these dressed coordinates should not be
introduced arbitrarilly. Since our problem is linear, we will require a linear
transformation between the normal and {\it dressed} coordinates (different from the 
transformation (\ref{transf}) linking the normal to the bare coordinates). Also, we demand 
the physical condition of vacuum stability.  We assume that at some given time ($t=0$) the 
system is described by {\it dressed} states, whose wavefunctions are defined by, 
\begin{equation} 
\psi_{\kappa_{0} \kappa_{1}...}(q^{\prime},0)=\prod_{\mu=0}^N
\left[\sqrt{\frac{2^{\kappa_{\mu}}}{\kappa_{\mu}}!)} 
H_{\kappa_{\mu}} (\sqrt{\bar{\omega}_{\mu}} 
q^{\prime}_{\mu})\right]\Gamma_{0}(q')\;,  
\label{ortovestidas1} 
\end{equation}
where $q^{\prime}_{\mu}=(q^{\prime}_{0}, q^{\prime}_{i})$, 
$\bar{\omega}_{\mu}=(\bar{\omega}, \omega_{i})$ and $\Gamma_{0}(q')$ 
has the same functional dependence on the variables $q'$ as 
 the ground state eigenfunction $\Gamma_0(Q)$ 
has on the variables $Q$, $i.e.$, can be obtained from $\Gamma_{0}(Q)$ by replacing $Q_r$ and $\Omega_r$ by the dressed variables  $q_\mu'$ and $\bar{\omega}_\mu$. 
The dressed states given by Eq. (\ref{ortovestidas1}) describe the dressed particle
in its $k_0$-th excited level in the precensse of $k_i$ dressed  field quanta 
(photons in the case of electromagnetic interactions) of
frequencies $\omega_i$.
Note that the above wavefunctions will 
evolve in time in a more complicated form than the unitary evolution of the eigenstates 
(\ref{autofuncoes}), since these wavefunctions are not eigenstates of the diagonal Hamiltonian 
(\ref{Hamiltoniana}).

In order to satisfy the physical condition of vacuum stability (invariance under a tranformation from normal to {\it dressed} coordinates) we  remember that the the ground 
state eigenfunction of the system has the form,
\begin{equation}
\Gamma_{0}(Q)\propto {\rm e}^{-\frac{1}{2\hbar}
\sum_{r=0}^N\Omega_r Q_r^2}\;,
\label{ad1}
\end{equation}
and we require that the ground state in terms of the {\it dressed} coordinates 
should have the form
\begin{equation}
\Gamma_{0}(q')\propto {\rm e}^{-\frac{1}{2\hbar}
\sum_{\mu=0}^N\bar{\omega}_\mu (q_\mu')^2}\;.
\label{ad2}
\end{equation}
From Eqs. (\ref{ad1}) and (\ref{ad2}) it can be seen that the vacuum invariance
requirement is satisfied if we define the {\it dressed} coordinates by,
\be
\sqrt{\bar{\omega}_\mu}q_\mu'=\sum_{r=0}^Nt_\mu^{r}\sqrt{\Omega_r}Q_r\;.
\label{eq11}
\ee

These {\it dressed} coordinates are new {\it collective} coordinates, different from 
the bare coordinates $q_\mu$ describing the bare particle and the 
free field modes, and also from the normal (collective) coordinates $\{Q_{r}\}$. 
Indeed, using Eq. (\ref{transf}) in Eq. (\ref{eq11}) the
{\it dressed} coordinates can be written in terms of the bare coordinates as
\begin{equation} 
q^{\prime}_{\mu}=\sum_{\nu=0}^N\alpha_{\mu \nu}q_{\nu}\;,
\;\;\;\;\;\alpha_{\mu \nu}=
\frac{1}{\sqrt{\bar{\omega}_{\mu}}}
\sum_{r=0}^Nt_{\mu}^{r}t_{\nu}^{r}\sqrt{\Omega_{r}}. 
\label{qvestidas3} 
\end{equation}

Our {\it dressed} states, given by Eq.(\ref{ortovestidas1}), are {\it collective} but 
{\it non} {\it stable} states (only the the lowest energy state, the invariant vacuum, 
is stable), linear combinations of the (stable) eigensatates 
(\ref{autofuncoes}) defined in terms of the normal coordinates. The coefficients of these 
combinations are given in Eq. (\ref{ortovestidas3}) below and explicit formulas for these 
coefficients for an interesting physical situation are given in Eq.(\ref{coeffN}). 
This gives  a complete and rigorous definition of our dressed states. Moreover, our 
dressed states have the interesting property of distributing the energy initially
in a particular dressed state, among itself and all other dressed states with precise
and well defined probability amplitudes \cite{adolfo1}. We {\it choose} these dressed
states as physically meaningful and we test successfully this hypothesis by studying 
the radiation process by an atom in a cavity. In both cases, of a very large or a very 
small cavity, our results are in agreement with experimental observations.

Using Eq. (\ref{eq11}) the functions given by Eq. (\ref{ortovestidas1}) can be expressed
in terms of the normal coordinates $Q_r$, but since the functions $\phi_{n_0,n_1,...}(Q)$ 
form a complete set of orthonormal functions, the functions $\psi_{k_0,k_1,...}(q')$can be written as linear combinations of the eigenfunctions of the coupled system 
(we take $t=0$ for the moment), 
\begin{equation} 
\psi_{\kappa_{0}\kappa_{1}...}(q^{\prime},0)=\sum_{n_{0}n_{1}...=0}^\infty
T_{\kappa_{0} \kappa_{1}...}^{n_{0}n_{1}...}\phi_{n_{0}n_{1}...}(Q,0)\;,  
\label{ortovestidas2} 
\end{equation} 
where the coefficients are given by,
\begin{equation} 
T_{\kappa_{0} \kappa_{1}...}^{n_{0}n_{1}...}=\int dQ\, \psi_{\kappa_{0}\kappa_{1}...}(q')
\phi_{n_{0}n_{1}...}(Q)\;,
\label{ortovestidas3} 
\end{equation} 
the integral extending over the whole $Q$-space. We consider the particular configuration 
$\psi$ in which only the dressed particle (atom) $q^{\prime}_{0}$ is in its $k$-th excited state, 
all other being in the ground state,
\begin{equation} 
\psi_{k00...}(q^{\prime},0)=(2^{-k}k!)^{-\frac{1}{2}} 
H_{k}(\sqrt{\bar{\omega}}q^{\prime}_{0})\Gamma_{0}(q')\;.  
\label{ortovestidas4} 
\end{equation} 
The above initial state represents a particle in the $k$-th excited level and no 
field quanta. 
The coefficients given by Eq. (\ref{ortovestidas3}) can be calculated in this case using
the theorem \cite{Ederlyi},
\begin{equation}
\frac{1}{k!}\left[\sum_{r}(t_{\mu}^{r})^{2}\right]^{\frac{k}{2}}
H_{k}(\frac{\sum_{r}t_{\mu}^{r}\sqrt{\Omega_{r}}Q_{r}}
{\sqrt{\sum_{r}(t_{\mu}^{r})^{2}}}) =\sum_{m_{0}+m_{1}+...=k}
\frac{(t_{\mu}^{0})^{m_{0}}(t_{\mu}^{1})^{m_{1}}...}{m_{0}!m_{1}!...}H_{m_{0}}
(\sqrt{\Omega_{0}}Q_{0})H_{m_{1}}
(\sqrt{\Omega_{1}}Q_{1})....  
\label{teorema Ederlyi} 
\end{equation}

Replacing Eq. (\ref{ortovestidas4}) in Eq. (\ref{ortovestidas3}) and using the theorem 
given by Eq. (\ref{teorema Ederlyi}), we get 
\begin{equation}
T_{k00...}^{n_{0}n_{1}...}=\sum_{n_0+n_1+...=k}(\frac{k!}{n_{0}!n_{1}!...})^{\frac{1}{2}}
(t_{0}^{0})^{n_{0}}(t_{0}^{1})^{n_{1}}...\;.  
\label{coeffN} 
\end{equation} 
Now we would like to compute the probability amplitude that at time $t$ the atom still
remain in the $k$-th excited level, that we denote by ${\mathcal A}_k(t)$, and given by
\be
{\mathcal A}_k(t)=\left(~\!\psi_{k00..}(q',0),~\!\psi_{k00...}(q',t)\!~\right)\;,
\label{la10}
\ee
where $\psi_{k00...}(q',t)$ can be obtained from Eq. (\ref{ortovestidas2}),
\be
\psi_{k00...}(q',t)=\sum_{n_0n_1...=0}^\infty
T_{k00...}^{n_0n_1...}e^{-iE_{n_0n_1...}t}\psi_{n_0n_1...}(Q,0)\;.
\label{la11}
\ee
Replacing Eq. (\ref{la11}) in Eq. (\ref{la10}) we get
\be
{\mathcal A}_k(t)=\sum_{n_0,n_1,...=0^\infty}\left(T_{k00...}^{n_0n_1...}\right)^2
e^{-iE_{n_0n_1...}}\;.
\label{la12}
\ee
Taking $k=1$ in Eq. (\ref{coeffN}) and replacing in Eq. (\ref{la12}), we obtain
for the probability amplitude that at time $t$ the atom still remain in the
first excited level,
\be
f^{00}(t)=\sum_{r=0}^N\left(t_0^r\right)^2 e^{-i\Omega_r t}\;,
\label{la13}
\ee
where we do not include a global phase factor that will not contribuite to the probability
$|f^{00}(t)|^2$ and we have changed the notation $f^{00}(t)={\mathcal A}_1(t)$.
In the case of a very large cavity ($L\rightarrow \infty$) the probability amplitude
that the dressed particle  be still excited at time $t$ can be obtained using 
Eq. (\ref{la8}) to transform the discrete sum given by Eq. (\ref{la13}) into an integral,
\begin{equation}
f^{00}(t)=2g\int_{0}^{\infty}d\Omega\frac{\Omega^{2}e^{-i\Omega t}} 
{(\Omega^{2}-\bar{\omega}^{2})^{2}+\pi^{2}g^{2}\Omega^{2}}\;.  
\label{f00}
\end{equation}
For large $t$ ($t>> \frac{1}{\bar{\omega}}$), but for in principle arbitrary coupling $g$, 
we obtain for the probability of finding the atom still excited at time $t$, the result 
\cite{adolfo3},  
\be
|f^{00}(t)|^2=\left(1+\frac{\pi^2g^2}{4\kappa^2}\right)e^{-\pi gt}
-e^{-\frac{\pi gt}{2}}\left[\frac{8g}{\bar{\omega}^4t^3}\left(\sin \kappa t+
\frac{\pi g}{2\kappa}\cos \kappa t\right)\right]
+\frac{16g^2}{\bar{\omega}^8t^6}\;,
\label{|f00|2} 
\end{equation} 
where $\kappa=\sqrt{\bar{\omega}^{2}-\frac{\pi^{2}g^{2}}{4}}$. 
In the above expression the approximation $t>>\frac{1}{\bar{\omega}}$ plays 
a role only in the two last terms, due to the difficulties to evaluate 
exactly the integral in Eq. (\ref{f00}) along the imaginary axis 
using Cauchy's theorem. The first term comes from the residue at 
$\Omega=\kappa+i\frac{\pi g}{2}$ and would be the same if we have done an 
exact calculation. If we consider in eq. (\ref{|f00|2}) $g<<\bar{\omega}$, which corresponds in electromagnetic theory to the fact that the fine structure constant $\alpha$ is small 
compared to unity (for explicit calculations we take below $g=\alpha\bar{\omega}$), 
we obtain the well known perturbative exponential decay law. For arbitrary time and
for arbitrary ratio $g/\bar{\omega}$ we can integrate Eq. (\ref{f00}) numerically and
we obtain an exponential like decay law.

Remark that in the limit of a very large cavity, we can express the dressed coordinates $%
q^{\prime}_{\mu}$ in terms of the bare ones, $q_{\mu}$ in the limit $%
L\rightarrow \infty$, 
\begin{equation}
q^{\prime}_{0}=A_{00}(\bar{\omega},g)q_{0},  \label{q0'q0}
\end{equation}
\begin{equation}
q^{\prime}_{i}=q_{i},  
\label{qi'qi}
\end{equation}
where $A_{00}(\bar{\omega},g)$, is given by.
\begin{equation}
 A_{00}(\bar{\omega},g),\frac{1}{\sqrt{\bar{\omega}}}%
\int_{0}^{\infty} \frac{2g\Omega^{2}\sqrt{\Omega} d\Omega}{(\Omega^{2}-\bar{%
\omega}^{2})^{2}+\pi^{2}g^{2}\Omega^{2}}.
\label{alfa00}
\end{equation}
It is interesting to compare Eqs.(\ref{qvestidas3}) with Eqs.(\ref{q0'q0}), (
\ref{qi'qi}). In the case of Eqs.(\ref{qvestidas3}) for finite $R$, the
coordinates $q^{\prime}_{0}$ and $\{q^{\prime}_{i}\}$ are all dressed, in
the sense that they are all collective, both the field modes and the
particle can not be separated in this language. In the limit $
R\rightarrow \infty$, Eqs.(\ref{q0'q0}) and (\ref{qi'qi}) tells us that the
coordinate $q^{\prime}_{0}$ describes the particle modified by
the presence of the field in a indissoluble way, the particle  
is always dressed by the field. On the other side, the dressed harmonic
modes of the field, described by the coordinates $q^{\prime}_{i}$ are
identical to the bare field modes, in other words, the field keeps in the
limit $R\rightarrow \infty$ its proper identity, while the particle  
 is always accompanied by a cloud of field quanta. Therefore we
identify the coordinate $q^{\prime}_{0}$ as the coordinate describing the
particle dressed by its proper field, being the whole system
divided into dressed particle and field, without appeal to the concept of
interaction between them, the interaction being absorbed in the dressing
cloud of the particle.
In the next section we will consider a partcle (an atom) confined in a small cavity. By "small" 
it is understood a cavity having dimensions much lower than macroscopic dimensions, 
but still much larger than atomic dimensions. In this case the concepts of dressed 
atom and of dressed field can be taken as the physically meaningful ones to describe the atom 
radiation process\,\cite{note}.

\section{The confined system}

Let us now consider the particle (an atom) placed in the center of the cavity in the case of  
a  small diameter $L$, $i.e.$ that satisfies the condition of 
being much smaller than the coherence lenght, $L<<2c/g$\,\cite{note}.  
To obtain the eigenfrequencies spectrum, we remark that 
from an analysis of Eq. (\ref{eigenfrequencies2}) it can be seen  
that in the case of a small values of $L$,  its  solutions are near the frequency values corresponding to the 
asymptots of the curve
$\cot(\frac{L\Omega}{2})$, which correspond to the field modes $\omega_{i}=2\pi i/L$.
The smallest solution departs more from
the first asymptot than the other larger solutions depart from 
their respective nearest asymptot. As  
we take larger and larger solutions, they are nearer and nearer 
to the values corresponding to the 
asymptots. For instance, for a value of $L$ of the order of $2\times 10^{-2}m$ 
and $\bar{\omega}\sim 10^{10}/s$, only 
the lowest eigenfrequency $\Omega_{0}$ is significantly different from the field frequency 
corresponding to the first asymptot, all the other eigenfrequencies $\Omega_{k},~
k=1,2,...$ being very close to the field modes $2\pi k/L$. For higher values of 
$\bar{\omega}$ (and lower values of $L$) the differences between the eigenfrequencies 
and the field modes frequencies are still smaller.
Thus to solve Eq. (\ref{eigenfrequencies2}) for the larger eigenfrequencies 
we expand the 
function $\cot(\frac{L\Omega}{2})$ around the values corresponding 
to the asymptots. Writing,
\begin{equation}
\Omega_k=\frac{2\pi }{L}(k+\epsilon_k),~~~k=1,2,..
\label{others}
\end{equation} 
with $0<\epsilon_{k}<1$, and substituting in Eq. (\ref{eigenfrequencies2}) we get,
\begin{equation}
\cot(\pi \epsilon_k)=\frac{c}{gL}(k+\epsilon_k) +\frac{1}{(k+\epsilon_k)}
(1-\frac{\bar{\omega}^{2}L}{2\pi g})\;.  
\label{eigen2} 
\end{equation}
But since for a small value of $L$ every $\epsilon_k$ is  much smaller than $1$, 
Eq. ({\ref{eigen2}) can be linearized in  $\epsilon_k$, giving, 
\begin{equation}
\epsilon_k=\frac{4\pi g  L k}{2(4\pi^{2} k^{2}-\bar{\omega}^{2}L^{2})}.
\label{linear}
\end{equation}
Eqs. (\ref{others}) and (\ref{linear}) give approximate solutions for the eigenfrequencies 
$\Omega_{k},~k=1,2...$.
To solve Eq. (\ref{eigenfrequencies2}) with respect to the lowest eigenfrequency 
$\Omega_{0}$, let us assume that it satisfies the condition 
$\Omega_{0}L/2<<1$ (we will see below that this condition is compatible with the 
condition of a small $L$ as defined above). 
Inserting the condition $\Omega_{0}L/2<<1$ in Eq. (\ref{eigenfrequencies2})
and keeping up to quadratic terms in $\Omega$ we obtain the solution for the lowest 
eigenfrequency  $\Omega_{0}$,
\begin{equation}
\Omega_0=\frac{\bar{\omega}}{\sqrt{1+\frac{\pi gL}{2c}}}.
\label{firsts}
\end{equation}
Consistency between Eq. (\ref{firsts}) and the condition  $\Omega_{0}L/2<<1$ gives 
a condition on $L$,
\begin{equation}
L\ll 
\frac{2c}{g}\frac{\pi}{2}\left[\frac{g}{\bar{\omega}}\right)^2
\left(1+\sqrt{ 1+\left(\frac{2\bar{\omega}}{\pi g}\right)^2}~\right].
\label{rsmall}
\end{equation}

{\bf a) Weak coupling}\\

Let us particularize our model to situations where interactions of electromagnetic type 
are involved. In this case,  
let us define the coupling constant $g$ to be such that $g=\bar{\omega}\alpha$, where 
$\alpha$ is the fine structure constant, $\alpha=1/137$. Then  
the factor multiplying $2/g$ Eq. (\ref{rsmall}) is $\sim 0.07$ and the condition $L\ll 2/g$ 
is replaced by a more restrictive one, $L\ll 0.07(2/g)$. For a typical infrared frequency, 
for instance $\bar{\omega}\sim 2,0\times 10^{11}/s$, our calculations are valid for a value of $L$,   
$L\ll 10^{-3}m$\,\cite{note}.
From Eqs.(\ref{t0r2}) and using the above expressions 
for the eigenfrequencies for small $L$, we obtain the matrix elements,
\begin{equation}
(t_0^0)^2\approx 1-\frac{\pi g L}{2}\;,~~~(t_0^k)^2\approx \frac{g L}{\pi  k^2}\;.
\label{too}
\end{equation}
To obtain the above equations we have neglected the corrective term $\epsilon_{k}$,
from the expressions for the eigenfrequencies $\Omega_{k}$.   Nevertheless, corrections in $\epsilon_{k}$ should be included in the expressions for
the matrix elements $t_{k}^{k}$, in order to avoid 
spurious singularities due to our approximation.
Let us consider the situation where the dressed atom is initially 
in its first excited level. Then from Eq. (\ref{la13}) we obtain the probability that 
it will still be excited after a ellapsed time $t$, 
\begin{equation}
|f^{00}(t)|^{2}=
(t_{0}^{0})^{4}+2\sum_{k=1}^{\infty}(t_{0}^{0})^{2}(t_{0}^{k})^{2}
\cos(\Omega_{k}-\Omega_{0})t+
\sum_{k,l=1}^{\infty}(t_{0}^{k})^{2}
(t_{0}^{l})^{2}\cos(\Omega_{k}-\Omega_{l})t\;,
\label{|f00R|2}
\end{equation}
or, using Eqs.(\ref{too}) in Eq.({\ref{|f00R|2}), 
\begin{equation}
|f^{00}(t)|^{2}\approx 1-\pi \delta+4(\frac{\delta}{\pi}-\delta^{2})
\sum_{k=1}^{\infty}\frac{1}{k^{2}}\cos(\Omega_{k}-\Omega_{0})t+
\pi^{2}\delta^{2}+\frac{4}{\pi^{2}}\delta^{2}\sum_{k,l=1}^{\infty}
\frac{1}{k^{2}l^{2}}\cos (\Omega_{k}-\Omega_{l})t,
\label{f002}
\end{equation} 
where we have introduced the dimensionless parameter $\delta=Lg/2\!~\ll 1$, corresponding
to a small value of $L$ and we remember that the eigenfrequencies are given by 
Eqs. (\ref{others}) and (\ref{linear}). 
As time goes on, the probability that the mechanical oscillator be excited  attains 
periodically a minimum value which has a lower bound given by,
\begin{equation}
\mathrm{Min}(|f^{00}(t)|^{2})=1-\frac{5\pi}{3}\delta+\frac{14\pi^{2}}{9}\delta^{2}.
\label{min}
\end{equation}
For a frequency $\bar{\omega}$ of the order $\bar{\omega}\sim 4.00\times 10^{14}/s$ 
(in the red visible), which corresponds to $\delta\sim 0.005$ and 
$L\sim 1.0\times 10^{-6}m$\,\cite{note}, we see from Eq.(\ref{min}) that the probability
that the atom be at any time excited will never fall below a value 
$\sim 0.97$, or a decay probability that is never higher that a value $\sim 0.03$.  
It is interesting to compare this result with experimental observations in 
\cite{Hulet,Haroche3}, where stability is found for atoms emiting in the visible range 
placed between two parallel mirrors a distance $L=1.1\times 10^{-6}m$ apart from one 
another. For lower frequencies the value of the spacing $L$ ensuring quasi-stability 
of the same order as above for the excited atom, may be considerably 
larger. For instance, for $\bar{\omega}$ in a typical microwave value, $\bar{\omega}\sim 2,00\times 10^{10}/s$ and taking also $\delta \sim 0.005$,  
the probability that the atom remain in the first excited level at any
time would be larger than a value of the order of $97\%$,  
for a value of $L$, $L\sim 2.0\times 10^{-2}m$. The probability that the   
mechanical oscillator remain excited as time goes on, oscillates with time 
between a maximum and a minimum values and never departs 
significantly from the situation of stability  in the excited state.
Indeed  for an emission frequency  $\bar{\omega}\sim 4.00\times 10^{14}/s$ (in the red visible) 
considered above and $L\sim 1.0\times 10^{-6}m$, the period of oscillation between the minimum
and maximum values of the probability that the dressed mechanical oscillator be excited,  
is of the order of $\frac{1}{12}\times 10^{-14}s$, while for $\bar{\omega}\sim 2,00\times 10^{10}/s$, 
and $L\sim 2.8\times 10^{-2}m$, the period is of the order of $\frac{1.4}{6}\times 10^{-10}s$.\

{\bf b) Strong coupling}\\

In this case it can be seen from Eqs (\ref{linear}), (\ref{firsts}) and 
(\ref{rsmall}) that $\Omega_{0}\approx \bar{\omega}$  and we 
obtain from Eq. (\ref{t0r2}),
\begin{equation}
(t_0^0)^2\approx \frac{1}{1+\pi \delta/2};\;\;(t_0^k)^2\approx \frac{g L}{\pi  k^2}\;.
\label{toos}
\end{equation}
Using Eqs.(\ref{toos}) in Eq.({\ref{|f00R|2}), we obtain for the probability that the 
excited particle be still at the first energy level at time $t$, the expression,
\begin{equation}
|f^{00}(t)|^{2}\approx \left(\frac{2}{2+\pi \delta}\right)^{2}+\frac{2}
{2+\pi \delta}\sum_{k=1}^{\infty}\frac{2\delta}{\pi k^{2}}
\cos(\Omega_{k}-\Omega_{0})t+
\frac{4}{\pi^{2}}\delta^{2}\sum_{k,l=1}^{\infty}\frac{1}{k^{2}l^{2}}
\cos (\Omega_{k}-\Omega_{l})t\;.
\label{f002s}
\end{equation}

We see from (\ref{f002s}) that as the system evolves in time, the probability that the 
particle still be excited, attains periodically a minimum value which has a lower bound given 
by,
\be
\mathrm{Min}(\delta)=\left(\frac{2}{2+\pi \delta}\right)^{2}-
\left(\frac{2}{2+\pi \delta}\right)\frac{\pi \delta}{3}-\frac{\pi^{2}\delta^{2}}{9}.
\label{mins}
\ee
The behaviour of the system in the strong coupling regime is completely different from 
weak coupling behaviour. The condition of positivity of (\ref{mins}) imposes for {\it fixed} 
values of $g$ and $\bar{\omega}$ an upper bound for the quantity $\delta$, $\delta_{max}$, 
which corresponds to an upper bound to the diameter $L$ of the cavity, $L_{max}$ 
(remember $\delta=Lg/2$). Values of $\delta$ larger than $\delta_{max}$, or equivalently, 
values of $L$ larger than $L_{max}$ are unphysical (correspond to negative probabilities) 
and should not be considered. These upper bounds are obtained from the solution with respect 
to $\delta$, of the inequality, $\mathrm{Min}(\delta)\geq 0$. We have 
$\mathrm{Min}(\delta)>0$ or $\mathrm{Min}(\delta)=0$, for respectively $\delta<\delta_{max}$ 
or $\delta =\delta_{max}$. The solution of the equation $\mathrm{Min}(\delta)=0$ gives
$\delta_{max}=-\frac{1}{\pi}+\frac{\sqrt{-2+3\sqrt{5}}}{\pi}\approx 0.37$.
For $\delta=\delta_{max}$, the minimal probability that the 
excited particle remain stable vanishes. 
We see comparing with the results of the previous subsection, that the behaviour of the system 
for {\it strong} coupling is rather different from its behaviour in the weak coupling regime. 
For appropriate cavity dimensions, which are of the same order  of those ensuring stability in 
the weak coupling regime, we ensure for strong coupling the complete decay of the particle to 
the ground state in a small ellapsed time.

\section{An anharmonic oscillator in a large cavity}

In this section we intend to generalize, making some appropriate approximations,to an {\it anharmonic} oscillator, the linear coupling to an environnement as 
it has been considered in the preceding sections. For more details see Ref.
\cite{adolfo4}. In this case, the whole system is described by the Hamiltonian,
\be
H[\lambda_r]=
H(p_{0},q_{0},\{p_{k},q_{k}\})+\sum_{r=0}^N\lambda_{r}{\cal T}^{(r)}_{\mu\nu\rho\sigma}
q_{\mu}q_{\nu}q_{\rho}q_{\sigma}\;,
\label{H1}
\ee 
where $H(p_{0},q_{0},\{p_{k},q_{k}\})$ is the bilinear Hamiltonian given by Eq.
(\ref{Hamiltoniana}) and ${\cal T}^{(r)}_{\mu\nu\rho\sigma}$ are some coefficients that will be defined below. In Eq. (\ref{H1}) summation over repeated greek 
labels is understood. We do not 
intend to go to higher orders in the perturbative series for the energy 
eigenstatates, we will remain at a first order correction in $\lambda_r$, 
and we will try to see what are the effects of the anharmonicity term, given 
by the last term of Eq. (\ref{H1}), on our previous results for the linear 
coupling with an environnement (field). We  notice that the anharmonicity term in 
Eq. (\ref{H1}) involves, independent quartic terms of the type $q_\mu^4$, 
$\mu=0,\{i\}$ (self-coupling of the bare oscillator and of the field modes), 
quartic terms coupling the oscillator to the field modes and also  the terms 
coupling the field modes among themselves. These terms are of the type 
$q_0q_i^3$, $q_0^2q_i^2$, $q_0^3q_i$, and of the type $q_i^nq_j^m$, with 
$n+m=4$, for the coupling between the field modes. We intend to start from 
the exact solutions we have found in the linear case, and investigate at
first order, how the quartic interaction characteristic of the anharmonicity 
changes the decay probabilities obtained in the linear case.

The introduction of the quartic interaction term in Eq.(\ref{H1})
changes the Hamiltonian in principal axis from Eq. (\ref{diagonal}) into
\begin{equation}
H[\lambda_r]=\frac{1}{2}\sum_{r=0}^{N}(P_{r}^{2}+\Omega_{r}^{2}Q_{r}^{2})+
\sum_{r=0}^N\lambda_r{\cal T}^{(r)}_{\mu\nu\rho\sigma}
t_\mu^{r_1}t_\nu^{r_2}t_\rho^{r_3}t_\sigma^{r_4}
Q_{r_1}Q_{r_2}Q_{r_3}Q_{r_4}\;,
\label{eixospri.2}
\end{equation}
where summation over the repeated indices $r_{1},r_{2},r_{3},r_{4}$ is 
understood. In order to have a specific quartic interaction, we make a 
choice for the coefficients ${\cal T}_{\mu\nu\rho\sigma}^{(r)}$ in the above 
equation,
\be
{\cal T}^{(r)}_{\mu\nu\rho\sigma}=t_{\mu}^rt_{\nu}^rt_{\rho}^rt_{\sigma}^r\;,
\label{T1}
\ee
which replaced in Eq. (\ref{eixospri.2}) and using the orthonormality of the 
matrix $t_\mu^r$ gives the Hamiltonian in principal axis,
\be
H[\lambda_r]=\frac{1}{2}\sum_{r=0}^{N}(P_{r}^{2}+\Omega_{r}^{2}Q_{r}^{2})
+\sum_{r=0}^N\lambda_rQ_r^4\;.
\label{H2}
\ee
Performing a perturbative calculation in $\lambda_r$, we can obtain the 
first order correction to the energy of the state corresponding to
$n_0=n_1=....=1$ in Eq. (\ref{autoe}),
in such a way that the $\lambda_r$-corrected energy of such state can be written in the form,
\be
E[\lambda_r]=\sum_{r=0}^N\left(\Omega_{r}+\lambda_r e_{r}^{(1)}\right)\;,
\label{energ.1}
\ee
where we do not include the energy of the ground state for the reason explained after
Eq. (\ref{la13}). The first order correction $e_{r}^{(1)}$ in Eq. (\ref{energ.1})
is given by
\be
e_{r}^{(1)}=\frac{15}{4\Omega_r^2}\;.
\label{first}
\ee
For sufficiently small $\lambda_r$ we will have quasi-harmonic normal 
collective modes having frequencies $\Omega_{r}+\lambda_r e_{r}^{(1)}$. 
Accordingly we can describe {\it approximatelly} the system in terms of 
modified harmonic eigenstates, which can be written as a generalization of 
the exact eigenstates (\ref{autofuncoes}), replacing the eigenfrequencies 
$\Omega_{r}$ by the $\lambda$-corrected values 
$\Omega_{r}+\lambda_{r}e_{r}^{(1)}$,
\begin{equation} 
\phi_{n_{0}n_{1}n_{2}...}(Q,t;\{\lambda_{s}\})=
\prod_{s=0}^N\left[\sqrt{\frac{2^{n_s}}{n_s!}}H_{n_{s}}(\sqrt{\frac{ 
\Omega_{s}}{\hbar}}Q_{s})\right]
\Gamma_{0}(Q)e^{-i\sum_{s}n_s\left(\Omega_{s}+\lambda_s e_{s}^{(1)}\right)t}\;. 
\label{autofuncoes1} 
\end{equation}
From the modified harmonic eigenstates (\ref{autofuncoes1}), we can follow 
analogous steps as in the harmonic case  to study the 
$\{\lambda_r\}$-corrected evolution of a dressed particle, generalizing Eq. (\ref{la13}),
obtaining
\be
f^{0 0}(t;\{\lambda_r\})=
\sum_{r=0}^N(t_{0}^{r})^2e^{-i\left(\Omega_{r}+\lambda_r e_{r}^{(1)}\right)t}\;.
\label{ortovestidas8} 
\ee

For a very large cavity, $L\to\infty$, using Eq. (\ref{la8}), we obtain we obtain 
the continuum limit,
\begin{equation}
f^{00}(t;\{\lambda_\Omega\})=2g\int_{0}^{\infty}d\Omega\frac{\Omega^{2}e^{-i\left(\Omega+
\frac{15}{4\Omega^2}\lambda_\Omega\right) t}} 
{(\Omega^{2}-\bar{\omega}^{2})^{2}+\pi^{2}g^{2}\Omega^{2}}.  
\label{f00(1)}
\end{equation}
From dimensional arguments, we can choose $\lambda_\Omega=\lambda\Omega^3$, where $\lambda$ is a dimensionless small fixed constant. With this choice, 
after expanding in powers of $\lambda$ the exponential in Eq. (\ref{f00(1)}), 
we obtain to first order in $\lambda$ the amplitude,
\be
f^{00}(t;\lambda)=f^{00}(t)+\frac{15\lambda t}{4}\frac{\partial}{\partial t} 
f^{00}(t)\;,
\label{f00(2)}
\ee
where $f^{00}(t)$ is the probability amplitude, for the harmonic problem, that
the particle be after a time $t$, still in the first excited level. 
From Eq. (\ref{f00(2)}) we can obtain at order $\lambda$, the probability 
that the particle remains in the first excited state at time $t$,
\be
|f^{00}(t;\lambda)|^2=|f^{00}(t)|^2+\frac{15\lambda t}{4}\frac{\partial}{\partial t}
|f^{00}(t)|^2\;.
\label{f00(4)}
\ee

We know that $|f^{00}(t)|^2$ is a decreasing function of $t$, what means that
the derivative of this function with respect to $t$ is negative. Therefore, 
since $\lambda>0$, we conclude from Eq. (\ref{f00(4)}) that 
$|f^{00}(t;\lambda)|^2$ is smaller than the harmonic probability
$|f^{00}(t)|^2$. 
In Fig. 1 we plot on the same scale the $\lambda$-corrected probability 
(\ref{f00(4)}) and the harmonic probability given by Eq. (\ref{|f00|2}), for 
$\bar{\omega}=4.0\times 10^{14}/s$ and $g=\alpha \bar{\omega}$, where 
$\alpha$ is the fine structure constant, $\alpha=1/137$ and the time is rescaled 
as  $t\times 10^{-13}s$. The solid line is the harmonic probability 
(\ref{|f00|2}) and the dashed line is the $\lambda$-corrected probability 
(\ref{f00(4)}), for $\lambda=1/50$. We see clearly the enhancement of the 
particle decay induced by the quartic interaction.

\begin{figure}[c] 

\epsfysize=6cm  

{\centerline{\epsfbox{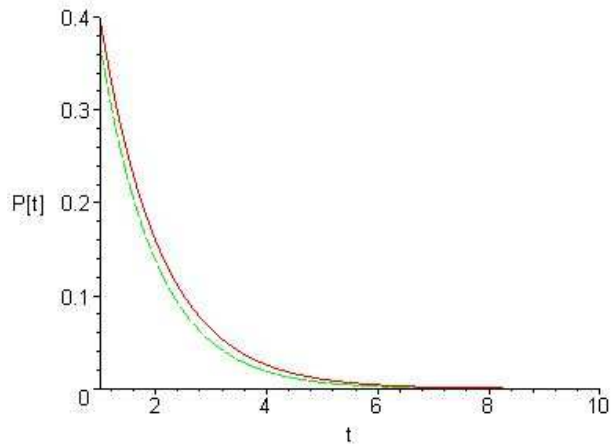}}}
\caption{Plot on the vertical axis, 
commonly named $P[t]$, of the $\lambda$-corrected probability (\ref{f00(4)}) 
(dashed line) and the harmonic probability  Eq.(\ref{|f00|2}) (solid line), 
for $\bar{\omega}=4.0\times 10^{14}/s$ and $g=\alpha \bar{\omega}$. $\alpha$ 
is the fine structure constant, $\alpha=1/137$ and time in units of 
$10^{-13}s$}
\end{figure}

\section{Conclusions}

In this paper we have analysed a linearized  version of an
particle-environment system and we have tried to give the more exact
and rigorous treatment we could to the problem. We have adopted a general
physicist's point of view, in the sense that we have renounced to approach very
closely to the real behaviour of a complicated non-linear system, as a quark-gluons 
coupled system, to study
instead a linear model. As a counterpart, an exact solution has been
possible.

We have presented  a summary of mathematical results previously 
obtained, to describe an {\it ohmic} quantum system consisting 
of a particle (in the larger sense of a "material body", an atom or a Brownian particle
for instance) coupled to an environment modeled by non-interacting oscillators. 
We have used  a formalism 
({\it dressed} coordinates and {\it dressed} states) that allows a non-perturbative 
approach to the 
 time evolution of the system, in rather different situations as  confinement  
 or in free space. 
 For finite $R$, our
system could be seen as a simplified linear  model to confined quarks and 
gluons inside a hadron. In  this case all {\it dressed} coordinates are effectively 
 dressed, in
the sense that they are all collective, both the field modes and the
particle can not be separated in this language  [see 
Eq.(\ref{qvestidas3})]. Of course the normal 
coordinates are also collective, but 
they correspond to stable eigenstates, no change in time exists for them. If we ascribe 
physical meaning to our dressed coordinates and states in the 
context of our model, matter and gauge quanta 
inside hadrons, can not be individualized as "quarks" 
and "gluons", all we have is a kind of quark-gluon "magma". Since quarks and gluons are permanently 
confined, we could roughly think that, in the context of the model studied here, quarks and 
gluons should not really exist, they would be an artifact of perturbation theory.

In the limit 
$R\rightarrow \infty$,  we get the result that 
  the dressed coordinate associated to the particle describes the particle modified by
the presence of the field in a indissoluble way, the particle 
is always dressed by the field. On the other side, the dressed harmonic
modes of the field, are in the limit $R\rightarrow \infty$
identical to the bare field modes, in other words, the field keeps in the free space 
limit its own identity, while the particle  
 is always accompanied by a cloud of field quanta [see Eqs.(\ref{q0'q0}) and (\ref{qi'qi})].

The study on the behaviour of particles
(for instance atoms in the harmonic approximation) confined in small cavities, shows that it is completelly different from the behaviour  
 in free space. We have implicitly assumed in this study that 
a small cavity is still much larger than atomic dimensions (which is indeed the case for
the experimental situation compared to our results, 
corresponding to a cavity diameter of $\sim 1\mu$), in such a way that the dressed 
particle could be a good approximation to the atom inside the cavity.  
In the first case the time evolution is very 
sensitive to the presence of boundaries, a fact that has been pointed out since a long 
time ago in the literature (\cite{Morawitz}, \cite{Milonni}, \cite{Kleppner}). Our 
{\it dressed} states approach gives an unified description for the dressing of 
a charged particle by the field modes and the time evolution in a cavity of arbitrary size, 
which includes microcavities and very large cavities (free space). 
If we assume that our dressed particle  
is a good representation for an atom under certain circumstances, we recover here with our formalism the 
experimental observation that excited states of atoms in sufficiently small 
cavities are stable for weak coupling. 
In the weak coupling regime, we are able to give formulas for the probability of an 
atom to remain excited for an infinitely long time, provided it is placed in a cavity of 
appropriate size. 
For an interaction 
of electromagnetic type, for an emission frequency in the visible red, the size of such   
 cavity is in good agreement with experimental observations. 
Also, our approach gives an exact result for emission in free space, generalizing the well 
known exponential decay law. 
The behaviour of the system for 
{\it strong} coupling is rather different from 
its behaviour in the weak coupling regime. For appropriate 
cavity dimensions, which are of the same order  of those ensuring stability in the weak 
coupling regime,
we ensure for strong coupling the complete decay of the particle to the ground state in a 
small ellapsed time. 
One possible conclusion is that by changing conveniently the physical and geometric 
parameters (the emission frequency, the strength of the coupling and the size of the 
confining cavity) our formalism theoretically allows a control 
on the rate of emission and of  the energy storage capacity (perhaps information?) in the cavity. Depending on the strength of the coupling,
the emission probability ranges from a complete stability to a very
rapid decay.

The formalism that we presented is rather general an applies to all systems that can be
modeled by a Hamiltonian of the type given by Eq. (\ref{Hamiltoniana}). Applications to the termalization and decoherence problems will be reported elsewhere
\cite{adolfo5,adolfo6}.

\section{Acknowledgements}
This work received financial support from CNPq (Brazilian National Research Council).


\begin{thebibliography}{99}
\bibitem{note} Notice that in any case we will consider a cavity diameter much larger 
than atomic dimensions. For instance, a cavity diameter of $1\,\mu$ considered for an 
emission frequency of $\bar{\omega}\cong 4.0\times 10^{14}\,s^{-1}$ is of the order of 
$10^{4}$ times the Bohr diameter.
\end{thebibliography}
\end{document}